\documentstyle[psfig,aps,prl]{revtex}
\begin{document}
\title{Volume Elements of Monotone Metrics on the $n \times n$
Density Matrices as Densities-of-States for Thermodynamic Purposes. I}
\author{Paul B. Slater}
\address{ISBER, University of
California, Santa Barbara, CA 93106-2150\\
e-mail: slater@itp.ucsb.edu,
FAX: (805) 893-7995}

\date{\today}

\draft
\maketitle
\vskip -0.1cm

\begin{abstract}
Among the monotone metrics on the $(n^{2} -1)$-dimensional convex set
of $n \times n$ density matrices, as Petz and Sud\'ar have recently
elaborated, there are a {\it minimal} (Bures) and a {\it maximal} one. 
We examine the proposition that it is physically
meaningful to treat  the volume elements of these metrics
as densities-of-states for thermodynamic purposes.
In the $n=2$ (spin-${1 \over 2}$) case, use of the maximal monotone metric,
in fact, does lead to the adoption of the Langevin (and {\it not} the
Brillouin) function --- thus, completely
 conforming with a recent probabilistic argument
of Lavenda.
Brody and Hughston also arrived at the Langevin function in an analysis based
on the {\it Fubini-Study} metric.
It is a matter of some interest, however, that in the {\it first}
(subsequently  modified) 
 version of their
paper, they had reported a different result,
one  fully
consistent with the alternative use of the {\it minimal} monotone metric.
In this part I of our investigation, we first study scenarios involving
{\it partially} entangled spin-${1 \over 2}$ particles ($n = 4, 6,\ldots$),
and then a certain three-level extension of the two-level systems.
In part II, we examine, in full generality, and with some
limited analytical success, the cases
$n =3$ and 4.
\end{abstract}


\hspace{1.4cm} Keywords: quantum entanglement, quantum statistical
thermodynamics, canonical ensembles, Bures 

\hspace{3.2cm} metric, monotone metrics, Bayesian analyses, information gains,
 coupled spin-${1 \over 2}$ sytems,

\hspace{3.2cm}  spin-1 systems, Langevin function, Brillouin function

\vspace{-0.1cm}

\tableofcontents

\section{Introduction}
The Horodecki's have recently studied \cite{horo1,horo2}
 the relationship
between quantum entanglement and the Jaynes inference scheme  based on the 
maximization of entropy \cite{jaynes,balian,buzek}.
They found that the Jaynes methodology  ``can lead to problems with processing
of entanglement''.
 In this study, we also examine connections between the
entanglement of quantum systems and their statistical/thermodynamic
properties,
but
 from a rather different
viewpoint (cf. \cite{slaterbrill}).
A principal motivation for us in undertaking this work has been
the recent contention of
Brody and Hughston \cite{brod}
that the conventional (Jaynes)
 density-matrix approach to the canonical ensemble is
semiclassical in certain respects (since it eliminates the
 weighting of the quantum
phase space volume), along with their accompanying presentation of 
 an alternative ``quantum canonical
 ensemble'' based on the
metrical geometry of this space. In particular, Brody and Hughston
 analyzed the
case of a spin-${1 \over 2}$ particle in a heat bath and arrived (in the
{\it first} version \cite{brodun}  of their paper \cite{brod}) at the 
partition function,
\begin{equation} \label{part}
Q(\beta) = 2 \sqrt{\pi} \Gamma ({3 \over 2}) (\beta h)^{-1} I_{1} (\beta h)
= \pi  (\beta h)^{-1} I_{1}(\beta h).
\end{equation}
Here,  $\beta ={1 \over k T}$ is the inverse temperature
($T$ being temperature),
$k$ is Boltzmann's constant, $h$ is the magnetic field, and
 $I_{1}(x)$ is a particular
 modified (hyperbolic) Bessel function ($I_{\nu}(x), \nu =1$)
 of the first kind.
(Such functions often appear in the distribution of spherical and
directional random variables \cite{robert}.)
 From (\ref{part}), one can derive
the expectation  of the energy\cite[eq. (14)]{brodun},
\begin{equation} \label{energy}
 E = -{\partial \log{Q(\beta)} \over \partial \beta}
= -{h I_{2}(\beta h) \over I_{1}(\beta h)} = -\mu B
{I_{2}(\mu B/k T) \over I_{1} (\mu B/k T)},
\end{equation}
where
 $\mu$ is the particle's magnetic moment, and $B$ is the external magnetic
field strength, with $h \equiv \mu B$.
(Ratios of modified Bessel functions, such as appear in (\ref{energy}),
 play ``an important role in Bayesian
analysis'' \cite{robert}. In the limit $\beta \to 0$, the expected value of 
the energy (\ref{energy}) is 0, while the variance about the expected
value is, then, ${h^2 \over 4}$.) The {\it semiclassical} analogue
 of (\ref{energy}) is
the Brillouin function \cite{slaterbrill},
\begin{equation}
 E = -h {I_{{1 \over 2}}(\beta h)
 \over I_{-{1 \over 2}}(\beta h)}= -h \tanh{\beta h}.
\end{equation}

Not only have Brody and Hughston expressed certain reservations and
qualifications regarding the Brillouin function, so  has
Lavenda \cite[p. 193]{lav}. He argued that the ``Brillouin function has to
coincide with the first moment of the distribution [for a two-level system
having probabilities
 ${e^x \over e^x +e^{-x}}$ and ${e^{-x} \over e^x +e^{-x}}$, where
$x ={\mu B \over k T}$],  and this means
that the generating function is $Z(x) =\cosh{x}$. Now, it will be appreciated
that this function can not be written as a definite integral, such as
\begin{equation} \label{laveq}
Z(\beta) ={1 \over 2} \int_{-1}^{1} e^{\beta x} \mbox{d}x =
{\sinh{\beta} \over \beta} = ({\pi \over 2 \beta})^{{1 \over 2}}
I_{{1 \over 2}}(\beta),
\end{equation}
because the integral form for the hyperbolic Bessel function,
\begin{equation}
I_{\nu}(x) ={(x/2)^\nu \over \sqrt{\pi} \Gamma({1 \over 2} + \nu)}
\int_{-1}^{1} e^{\pm xt} \sin{t}^{\nu- {1 \over 2}} \mbox{d} t,
\end{equation}
exists only for $\nu > {1 \over 2}$. This means that $I_{-{1 \over 2}}(x)
= ({2 \over \pi x})^{{1 \over 2}}$ cannot be expressed in the above integral
form. Since the generating function cannot be derived as the Laplace transform
of a prior probability density, it casts serious doubts on the probabilistic
foundations of the Brillouin function. In other words, any putative
expression for the generating function must be compatible with the underlying
probabilistic structure; that is, it must be able to be represented as the
Laplace transform of a prior probability density'' (see also the further
detailed
remarks of Lavenda, to much the same effect \cite[pp. 20, 198]{lav}).
 Since the model yielding (\ref{part}) and (\ref{energy})
is based on the integral form for $I_{\nu}(x)$, for $\nu =1 > {1 \over 2}$,
 it clearly accords with the requirements of Lavenda.

Brody and Hughston \cite{brod}
 raised the possibility that the quantum canonical ensemble
could be distinguished from the conventional canonical ensemble by a suitable
measurement on a sufficiently {\it small} quantum mechanical system. In such
a case, they argued
 there would not seem to be any {\it a priori} reason for adopting
the semiclassical mixed state approximation, which allows random phases to be
averaged over.
Park and Band, in an extended series of papers \cite{band},
 expressed various qualms regarding the
conventional  (Jaynes) approach.
Park \cite{park1} himself later
 wrote that ``the details of quantum
thermodynamics are presently unknown'' and `` perhaps there is more to
the concept of thermodynamic equilibrium than can be captured in the canonical
density operator itself.''
Friedman and Shimony \cite{friedman}, in a classical rather than quantum
context, claim to have ``exhibited an anomaly in Jaynes' maximum entropy
prescription''.

The (initially presented) results  of Brody and Hughston \cite{brodun},
 that is, 
 (\ref{part}) and, implicitly, (\ref{energy}),  had, in fact, 
 been reported somewhat earlier by Slater \cite{slat1},
 along with parallel formulas for the {\it quaternionic} two-level systems,
in particular, the partition function (cf. (\ref{part})),
\begin{equation} \label{quaternion}
 Q(\beta) = 4 \sqrt{\pi} \Gamma({5
\over 2}) (\beta h)^{-2} I_{2}(\beta h) = 3 \pi (\beta h)^{-2} I_{2}(\beta
h) /4.
\end{equation}
This other
 analysis \cite{slat1}, similarly to \cite{brod},
 relied upon a metrical geometry, but the two
 approaches pursued appear, at least superficially, to be somewhat different.
The work of Brody and Hughston employed the {\it Fubini-Study}
 metric on the complex
projective space $CP^{n}$ (the space of rays --- which
 they regarded as the ``true 'state space'
of quantum mechanics''). The study of Slater, on the other hand, utilized
the {\it Bures} metric, which is defined on the space of density matrices
 \cite{hub1,hub2,braun1}.
However, Petz and Sud\'ar \cite{petz} have recently shown that the extension
of the Bures metric to the pure states is exactly the Fubini-Study metric,
and that, in point of fact, the Bures metric is the only {\it monotone}
metric which admits such an extension.
 (The Bures metric is the {\it minimal} monotone
metric \cite{petz}. The Bures distance --- known also as the
Hellinger or Kakutani inner product --- is defined as an inner product of
two probability measures on some measurable space  \cite{kakutani}. 
Therefore, it can be naturally defined on the space of pure states, namely,
$CP^{n}$.) So, these two approaches may be demonstrably
 fully consistent with
one another --- as their agreement in yielding
 the results (\ref{part}) and (\ref{energy})
might lead one to hypothesize. However, we must bear in mind that Brody and
Hughston were led, in the second version of their paper
\cite{brod}, to revise certain of
 their original conclusions \cite{brodun}, in the manner
indicated at the beginning of sec.~\ref{parallel}.
 (In their several recent joint papers dealing with
quantum statistical issues, Brody and Hughston
have chosen to
``emphasize the role of the space of {\it pure} quantum states, since in the
Hilbert space based classical-quantum statistical correspondence this is the
state space that arises as the immediate object of interest. In fact, the
space of density matrices has a very complicated structure, owing
essentially to the various levels of `degeneracy' a density matrix can 
possess, and the relation of these levels to one another'' \cite{brod2}.)

In the present communication, we follow in sec.~\ref{minimal}
 the specific line of argument of
 Slater \cite{slat1} (based on the Bures metric), with the objective of
developing ``quantum canonical ensembles'' for higher-dimensional situations 
than the (two-dimensional) one presented by a single spin-${1 \over 2}$
particle, previously analyzed. We examine several different scenarios for
 partially
 entangled spin-${1 \over 2}$ particles, and conduct an analysis of the same
nature in
sec.~\ref{s5} for a certain {\it three}-level extension of the two-level
systems, previously studied
 \cite{slat2}. Then, in sec.~\ref{parallel}, we reexamine the
scenarios studied in sec.~\ref{minimal}, but now with the use of the 
{\it maximal} rather than minimal monotone metric.
In an appendix, we report information gains from possible outcomes of joint
measurements of the partially entangled spin-${1 \over 2}$ particles.

We would also like to bring to the reader's attention, the second part
of this study
\cite{slaternew}, in which certain 
results (indicated in sec.~\ref{concluding},
 based on the {\it maximal} monotone metric, the subject 
of 
sec.~\ref{parallel}) of a surprisingly simple nature 
have been obtained for the most general
(unrestricted)
form of scenario for a spin-1 particle (describable by a $3 \times 3$
density matrix with eight free parameters). In \cite{slaternew}, we also
derived some results pertaining to the 
(yet unrealized) possibility
of extending the form of analysis taken there
 to general spin-${3 \over 2}$, or equivalently
{\it arbitrarily} 
coupled or entangled spin-${1 \over 2}$ systems (describable by a 
$4 \times 4$ density matrix with fifteen free parameters).
(We are confronted there with a challenging problem
of diagonalizing a $6 \times 6$ symmetric matrix.)

\section{Analyses Based on the {\it Minimal} Monotone 
 (Bures) Metric} \label{minimal}
 In general, the 
 $4 \times 4$ density
matrix ($\rho^{(a,b)}$) of a pair of  (arbitrarily entangled)
 spin-${1 \over 2}$ particles
($a,b$) can be written in the form (we adopt the
notation of \cite{chal}),
\begin{equation} \label{chal}
\rho^{(a,b)} = {1 \over 4} \{ I^{(a)} \otimes I^{(b)} +\xi^{(a)} \sigma^{(a)}
\otimes I^{(b} +I^{(a)} \otimes \xi^{(b)} \sigma^{(b)} +
\sum_{i,j=1}^{3} \zeta_{i j} \sigma_{i}^{(a)} \otimes \sigma_{j}^{(b)} \},
\end{equation}
where $I^{(a),(b)}$ and $\sigma^{(a),(b)}$ are Pauli matrices acting in the
space of particle $a$ and $b$, respectively.
The three-vectors $\xi^{(a),(b)}$, where $\xi^{(a)} = (\xi_{1}^{(a)},
\xi_{2}^{(a)},\xi_{3}^{(a)})$, correspond
 (in the case of photons) to the Stokes
vectors, while the parameters $\zeta_{i j}$ describe the interparticle
correlations. If the two particles are independent (nonentangled),
 then, $\zeta_{i j}
=\xi_{i}^{(a)} \xi_{j}^{(b)}$. In all the scenarios considered below, it is
assumed that particle $a$ is described by the same $2 \times 2$ density matrix
as particle $b$, that is, $\rho^{(a)} = \rho^{(b)}$, or equivalently,
$\xi^{(a)} = \xi^{(b)}$.

\subsection{Particles $a$ and $b$ are
 Polarized and Correlated with Respect to the Same
 Direction} \label{s1}
For the first of several scenarios to be examined,
 let us set twelve of the fifteen parameters
 in the expansion (\ref{chal}) {\it ab initio} to zero,
leaving only: (1) $\xi_{1}^{(b)}$,
 which we equate to $\xi_{1}^{(a)}$;  and (2)
$\zeta_{1 1}$. This corresponds to a situation in which the two particles
are unpolarized and uncorrelated except in one direction (associated with the
index
 ``1'') of  three underlying
orthogonal directions. Thus, we are concerned with a {\it doubly}-parameterized
set of $4 \times 4$ density matrices. These have the eigenvalues,
\begin{equation} \label{eigenvalues}
 {1 - \zeta_{1 1} \over 4},\quad {1- \zeta_{1 1} \over 4},\quad
{1 \over 4} (1 + 2 \xi_{1}^{(a)} +\zeta_{1 1})
,\quad  {1 \over 4} (1 - 2 \xi_{1}^{(a)}
+\zeta_{1 1}) 
\end{equation}
and corresponding  eigenvectors,
\begin{equation} \label{eigenvectors}
(0,0,1,0),\quad (0,1,0,0),\quad (1,0,0,0)), \quad (0,0,0,1)
\end{equation}
(Such a system is in a state of degeneracy if either $\zeta_{1 1}=1$ or
$\zeta_{1 1} = -1 \pm 2 \xi_{1}^{(a)}$, since at least one of the
eigenvalues is, then, zero. For the system to be in a pure state, we must
have $\zeta_{1 1}=1$ and $\xi^{(a)} = \pm 1$.)
We employed these results ((\ref{eigenvalues}), (\ref{eigenvectors}))
in the formula of H\"ubner \cite{hub1} for the Bures metric,
\begin{equation} \label{Bures}
d_{B} (\rho,\rho +\mbox{d} \rho)^{2} =\sum_{i, j = 1}^{n} {1 \over 2}
{ |<i | \mbox{d} \rho | j>|^{2} \over \lambda_{i} +\lambda_{j}},
\end{equation}
where $\lambda_{i}$ is the $i$-th eigenvalue and $<i|$ the corresponding
eigenvector of an $n \times n$ density matrix $\rho$.
For the case at hand, we have the result,
\begin{equation} \label{Bures2}
d_{B}(\rho, \rho +\mbox{d} \rho)^2 = g_{\xi_{1}^{(a)} \xi_{1}^{(a)}} \mbox{d}
{\xi_{1}^{(a)}}^{2} + g_{\xi_{1}^{(a)} \zeta_{1 1}} \mbox{d} \xi_{1}^{(a)} \mbox{d}
\zeta_{1 1} +g_{\zeta_{1 1} \zeta_{1 1}} \mbox{d} \zeta_{1 1}^{2},
\end{equation}
where
\begin{equation} \label{BB}
g_{\xi_{1}^{(a)} \xi_{1}^{(a)}} = {1 +\zeta_{1 1} \over 2
 ( -4 {\xi_{1}^{(a)}}^{2}
+(1 +\zeta_{1 1})^{2})},
\end{equation}
\begin{equation}
g_{\xi_{1}^{(a)} \zeta_{1 1}} = {\xi_{1}^{(a)} \over 4
 {\xi_{1}^{(a)}}^{2} - (1 +\zeta_{1 1})^{2}},
\end{equation}
and
\begin{equation} \label{BBB}
g_{\zeta_{1 1} \zeta_{1 1}} = {1 - 2 {\xi_{1}^{(a)}}^{2} +\zeta_{1 1} \over
4 (-1 +2 \xi_{1}^{(a)} - \zeta_{1 1}) (- 1 + \zeta_{1 1}) (1 +2 \xi_{1}^{(a)}
+ \zeta_{1 1})}.
\end{equation}
The corresponding volume element of the Bures metric is
(cf. \cite{horo3})
\begin{equation} \label{volume}
\sqrt{ g_{\xi_{1}^{(a)} \xi_{1}^{(a)}} g_{\zeta_{1 1} \zeta_{1 1}}
 -({g_{\xi_{1}^{(a)}
\zeta_{1 1}} \over 2})^{2}} = {1 \over 2 \sqrt{2}}
 \sqrt{ 1 \over (-1 + 2 \xi_{1}^{(a)} -\zeta_{1 1}) (-1 +\zeta_{1 1})
(1 + 2 \xi_{1}^{(a)} +\zeta_{1 1})}.
\end{equation}
If we integrate this volume element, first, over $\xi_{1}^{(a)}$ from
$-{1 \over 2} -{\zeta_{1 1} \over 2}$ to ${1 \over 2} +{\zeta_{1 1} \over 2}$
and, then, over $\zeta_{1 1}$ from -1 to 1, we obtain the result ${\pi
\over 2}$.
(These limits define the domain of feasible values of $\xi_{1}^{(a)}$ and
$\zeta_{1 1}$ --- which determine a triangular region --- for
 our doubly-parameterized density matrix.
Outside this region, not all the eigenvalues of $\rho^{(a,b)}$ lie between
0 and 1, as they must.)
 Dividing (\ref{volume}) by ${\pi \over 2}$, we obtain a (prior) probability
distribution \cite{slat7}
 over the  domain of these doubly-parameterized $4 \times 4$ density
matrices. Again, integrating the resultant probability distribution over
$\xi^{(a)}$, between the same limits as before,
 we obtain a univariate probability
distribution,
\begin{equation} \label{marginal}
i {\log{(1 + \zeta_{1 1}) - \log(-1 -\zeta_{1 1})} \over 2 \pi
 \sqrt{2}  \sqrt{1
-\zeta_{1 1}}},
\end{equation}
over the interval $\zeta_{1 1} \in [-1,1]$.
If we now regard (\ref{marginal}) as a (normalized)
 structure function or density-of-states
for thermodynamic purposes, multiply it by a Boltzmann factor,
$e^{-\beta h \zeta_{1 1}}$, and integrate over $\zeta_{1 1}$ from -1 to 1,
we obtain the partition function,
\begin{equation} \label{partcoupled}
Q(\beta) ={e^{-\beta h} \sqrt{\pi} \mbox{erfi}
 (\sqrt{2 \beta h})
\over 2 \sqrt{ 2 \beta h}},
\end{equation}
where erfi denotes the imaginary error function ${\mbox{erf}(iz) \over i}$.
(The error function $\mbox{erf}(z)$ is the integral of the Gaussian
distribution, that is,  ${2 \over \sqrt{\pi}} \int_{0}^{z} e^{-t^{2}} \mbox{d}
t$.)
For the expected value of the ``energy'' ($h \zeta_{1 1}$), we have, then,
\begin{equation} \label{energy2}
-{\partial \log{Q(\beta)} \over \partial \beta} =
{1 \over 2 \beta} + h  - {h \sqrt{2} e^{2 \beta h} \over
\mbox{erfi} (\sqrt{2 \beta h}) \sqrt{\pi \beta h}}.
\end{equation}
As $\beta \to 0$, this expected value approaches ${h \over 3}$, while the
variance about the expected value approaches ${16 h^2 \over 45}$.

If we choose in this scenario to first integrate the volume element
(\ref{volume}) over $\zeta_{1 1}$ (rather than $\xi_{1}^{(a)}$) from
$-1 -2 \xi_{1}^{(a)}$ to $-1 +2 \xi_{1}^{(a)}$,
then we obtain (where $K$ represents the complete elliptic integral of the
first kind),
\begin{equation}
{K({2 \xi_{1}^{(a)} \over \xi_{1}^{(a)} +1}) \over 2 \sqrt{-\xi_{1}^{(a)}
-1}},
\end{equation}
which can not be explicitly integrated over $\xi_{1}^{(a)}$
from -1 to  1 (nor if, first,  multiplied by a
 Boltzmann factor, $e^{-\beta h \xi_{1}^{(a)}}$).
\subsection{Particles $a$ and $b$ are Polarized and Uncorrelated
 in One Direction, Unpolarized and Correlated in An Orthogonal
Direction} \label{s2}
Let us modify the previous scenario (sec \ref{s1})
 slightly by now setting the formerly free parameter $\zeta_{1 1}$
to 0, and  letting $\zeta_{2 2}$ be free instead.
 We still maintain
$\xi_{1}^{(b)} = \xi_{1}^{(a)}$, with the remaining twelve
parameters once again set to zero.
This corresponds to a situation in which the particles $a$ and $b$ are
correlated in one particular direction (labeled ``2''), but unpolarized
in that direction.
The elements of the Bures metric are, then,
\begin{equation} \label{g1}
g_{\xi_{1}^{(a)} \xi_{1}^{(a)}} = {-1 +\zeta_{2 2}^2
\over 2 (-1 +4 {\xi_{1}^{(a)}}^2 +\zeta_{2 2}^2)},
\end{equation}
\begin{equation}
g_{\xi_{1}^{(a)} \zeta_{2 2}} = {\xi_{1}^{(a)} \zeta_{2 2}
\over 1 - 4 {\xi_{1}^{(a)}}^2 -\zeta_{2 2}^2},
\end{equation}
and
\begin{equation} \label{g3}
g_{\zeta_{2 2} \zeta_{2 2}} = {1 -\zeta_{2 2}^2 +2 {\xi_{1}^{(a)}}^2
(-2 +\zeta_{2 2}^2) \over 4 (-1 +\zeta_{2 2}^2) (-1 +4 {\xi_{1}^{(a)}}^2
+\zeta_{2 2}^2)}.
\end{equation}
The integrations of the corresponding volume element of the Bures metric
((\ref{g1})-(\ref{g3})) are now performed, first,  over $\xi_{1}^{(a)}$ from
$-{\sqrt{1-\zeta_{2 2}^2} \over 2}$ to ${\sqrt{1-\zeta_{2 2}^2} \over 2}$ and,
then, over $\zeta_{2 2}$ from -1 to 1. (The feasible values lie within
an ellipse, the equation of which is $ 4{ \xi_{1}^{(a)}}^2
+\zeta_{2 2}^2 =1$.) This gives us a result of
${\pi \over 2 \sqrt{2}}$, which we can use to obtain a normalized volume
element, that is, a (prior) probability distribution,
\begin{equation} \label{normvol}
{1 \over \pi \sqrt{1 -4 {\xi_{1}^{(a)}}^2 -\zeta_{2 2}^2}},
\end{equation}
over the domain of the doubly-parameterized $4 \times 4$ density matrices
for this scenario.
The univariate marginal distribution of (\ref{normvol}) over $\zeta_{2 2}
\in [-1,1]$ is simply uniform (${1 \over 2}$) --- which we take as our
(normalized) structure function. Applying the Boltzmann factor,
$e^{-\beta h \zeta_{2 2}}$ to it, gives us a partition function
 (cf. (\ref{laveq})),
\begin{equation} \label{lang1}
Q(\beta) = {\sinh{\beta h} \over \beta h} = ({\pi \over 2 \beta h})^{1 \over 2}
I_{1 \over 2} (\beta h),
\end{equation}
yielding  an expected value of the ``energy'' ($h \zeta_{2 2}$) equal to
\begin{equation} \label{lang2}
-{\partial \log{Q(\beta)} \over \partial \beta} =
-h {I_{{3 \over 2}} \over I_{{1 \over 2}}}= {1 \over \beta} - h
\coth{\beta h}.
\end{equation}
(In the limit $\beta \to 0$, this expected value approaches 0, while the
corresponding
variance  approaches ${h^2 \over 3}$.)
It should be noted that the results (\ref{lang1}) and
(\ref{lang2}) are formally equivalent to those
given by the Langevin model of paramagnetism
\cite{tusy}.
\subsection{Particles $a$ and $b$ are
 Polarized and Uncorrelated in One Direction, Unpolarized and
 Equally
Correlated in the Two Orthogonal Directions} \label{s2-3}
The only difference between this scenario and the previous one (sec.~\ref{s2})
 is that the
correlation $\zeta_{3 3}$ is not set to 0, but rather equated to $\zeta_{2 2}$.
(So, although the particles $a$ and $b$ are unpolarized in the third
direction, the outcomes of their individual
 spin measurements in this direction may
be correlated with one another.)
This leads to a more simple outcome. The normalized form of the Bures
volume element is, now, 
\begin{equation} \label{newprior}
{4 \over \pi^2 \sqrt{1 -4 {\xi_{1}^{(a)}}^2} \sqrt{1 - 4 \zeta_{2 2}^2}}.
\end{equation}
The domain of feasible values is the square defined by the lines
$\xi_{1}^{(a)} = \pm {1 \over 2}$ and $\zeta_{2 2} = \pm {1 \over 2}$.
(So, the bivariate probability distribution (\ref{newprior}) factors
into the product of two univariate probability distributions.)
We can, then, multiply (\ref{newprior}) by the  bivariate Boltzmann
factor $e^{-\beta_{\xi} h \xi_{1}^{(a)} -\beta_{\zeta} h \zeta_{2 2}}$ and
integrate over the square region to obtain the (product) partition function,
\begin{equation} \label{new2}
Q(\beta_{\xi},\beta_{\zeta}) = I_{0} ({\beta_{\xi} h \over 2})
I_{0} ({\beta_{\zeta} h \over 2}).
\end{equation}

We have also obtained  partition functions of the form
 $I_{0}({\beta h \over 2})$ in two quite different scenarios, in which we set
thirteen of the fifteen parameters in the expansion (\ref{chal}) to zero
 and, otherwise,
set the cross-correlation
 $\zeta_{2 1}$ equal to either $\zeta_{1 2}$ or to $- \zeta_{1 2}$.

\subsection{Particles $a$ and $b$ are
 Unpolarized, but Equally Correlated in Three Orthogonal
Directions} \label{s3}
In this scenario, we set the six components of the two vectors $\xi^{(a)}$
and $\xi^{(b)}$ to zero, so the particles $a$ and $b$ are assumed to be
 unpolarized
in each of the three orthogonal directions (and, thus, with respect to any
arbitrary orientation). We also fix $\zeta_{i j} = 0$ if
$i \neq j$, so there are no correlations between spin measurements
in different directions.
(So, in these respects, $a$ and $b$ are independent or nonentangled.)
Finally, we set $\zeta_{3 3}= \zeta_{2 2} =\zeta_{1 1}$, so correlations are
allowed between the measurements of $a$ and $b$
 in the same direction. Thus, we are concerned here, not with a
doubly-parameterized family as in the first two analyses, but with a
singly-parameterized one.

We obtain as our prior probability distribution
 over the feasible range $\zeta_{1 1}
\in [-1,1/3]$,
\begin{equation} \label{p1}
{\sqrt{3} \over \pi \sqrt{1 -3 \zeta_{1 1}} \sqrt{1+\zeta_{1 1}}},
\end{equation}
from which one obtains the partition function,
\begin{equation} \label{p2}
 Q(\beta) = e^{{\beta h \over 3}} I_{0}({2 \beta h \over 3}),
\end{equation}
and an expected value of the ``energy'' ($h \zeta_{1 1}$) of
\begin{equation} \label{p3}
- {\partial \log{Q(\beta)} \over \partial \beta} =
{h \over 3}  (-1 - {2 I_{1}({2 \beta h \over 3}) \over I_{0}({2 \beta h
\over 3})}).
\end{equation}
So, we again encounter a ratio of modified Bessel functions
(cf. (\ref{energy})) \cite{robert}. The value of (\ref{p3}) at $\beta = 0$
is $- {h \over 3}$ while that of the associated variance is
the square of this, that is ${2 h^2 \over 9}$.
 
\subsection{Three and More  Unpolarized Particles having Equal
Highest-Order Intradirectional
Correlations} \label{s4}
The first of several scenarios discussed in this section
 might be regarded as a {\it three}-particle ($a,b,c$)
 analogue
of the  two-particle one presented in the immediately preceding section
(\ref{s3}). For a general $8 \times 8$
density matrix ($\rho^{a,b,c}$) representing the joint state of the three
particles, we have an expansion analogous to (\ref{chal}). In this expansion,
we consider all the 63 parameters to equal 0, except for the three
($\zeta_{1 1 1},\zeta_{2 2 2},\zeta_{3 3 3}$) representing the highest-order
intradirectional correlations. We regard these three parameters as having a
common value, which is designated $\zeta_{1 1 1}$. In other words, there is a 
possibly nonzero correlation between the outcomes of spin measurements
in some fixed direction for the three particles.

The  normalized volume element of the corresponding Bures metric is, then,
\begin{equation} \label{unsucc}
{\sqrt{8  \zeta_{1 1 1}^2 - 3} \over E({8 \over 9})
 \sqrt{12 \zeta_{1 1 1}^2 - 4}},
\end{equation}
where $E$ represents the complete elliptic integral, and the range of
feasible values is
 $\zeta_{1 1 1} \in [-1/\sqrt{3},1/\sqrt{3}]$. However, no
  explicit formula for the partition function was found.

Let us continue this line of analysis to the {\it four}-particle case.
Now, there are 255 parameters in the expansion analogous to (\ref{chal}).
We set 252 of them to 0, and equate both of the correlations
$\zeta_{2 2 2 2}$ and $\zeta_{3 3 3 3}$ to $\zeta_{1 1 1 1}$, so again we
are concerned with a singly-parameterized family of density matrices.
The feasible range of $\zeta_{1 1 1 1}$ is
 the interval [-1/3,1]. We are able
to normalize the volume element of the Bures metric over this interval,
obtaining the probability distribution (cf. (\ref{p1})),
\begin{equation} \label{four1}
{\sqrt{3} \over \pi \sqrt{1- \zeta_{1 1 1 1}} 
 \sqrt{1 +3 \zeta_{1 1 1 1}}},
\end{equation}
and the partition function (cf. (\ref{p2}))
\begin{equation}
Q(\beta) = e^{- {\beta h \over 3}} I_{0}({2 \beta h \over 3}),
\end{equation}
giving
 an expected value of the ``energy'' ($h \zeta_{1 1 1 1}$) (cf. (\ref{p3})),
\begin{equation} \label{four3}
-{\partial \log{Q(\beta)} \over \partial \beta} =
{h \over 3}  (1 - {2 I_{1}({2 \beta h \over 3}) \over I_{0}({2 \beta h
\over 3})}).
\end{equation}
For $\beta = 0$, this expected value equals ${h \over 3}$, while the 
associated variance is ${2 h^2 \over 9}$.

For the {\it five}-particle analogue, the thermodynamic behavior was
precisely the same as for the three-particle case discussed just above
(\ref{unsucc}),
with the replacement of $\zeta_{1 1 1}$ by $\zeta_{1 1 1 1 1}$.
For the {\it six}-particle analogue, the same results
 ((\ref{p1})-(\ref{p3}))
were obtained as in the {\it two}-particle case of sec.~\ref{s3}
 (replacing $\zeta_{1 1}$
by $\zeta_{1 1 1 1 1 1}$).
\subsection{Three Unpolarized Particles having Equal {\it Second}
Highest-Order Intradirectional Correlations} \label{s10}
We modify the 
 scenarios of the previous section (\ref{s4}) by, now, requiring
the highest-order correlations to be 0, while equating all the second-order
correlations to each other, obtaining, thereby, the one free parameter.
 In the three particle case, there are nine such
correlations --- the assumed common value of which,
 we denote by $\zeta_{1 1 0}$.
The corresponding prior probability distribution over the feasible
range $\zeta_{1 1 0} \in [-1/3,1/3]$ is, then,
\begin{equation} \label{sd1}
{6 \over \pi (4 -36 \zeta_{1 1 0}^2)},
\end{equation}
yielding a partition function,
\begin{equation}
 Q(\beta) = I_{0}({\beta h \over 3}).
\end{equation}
The expected value of the ``energy'' ($h \zeta_{1 1 0}$) is,
\begin{equation} \label{sd3}
-{\partial \log{Q(\beta)} \over \partial \beta} =
-{h \over 3} {I_{1} ({\beta h \over 3}) \over I_{0} ({\beta h \over 3})}.
\end{equation}
For $\beta = 0$, this is equal to 0, with a corresponding variance of
${h^2 \over 18}$.

In the four-particle analogue, we have twelve second-highest order
correlations --- the assumed common value of which, denoted
$\zeta_{1 1 1 0}$, has a feasible range of $[-{1 \over 4 \sqrt{3}},{1 \over
4 \sqrt{3}}]$.
However, we have been unable to determine the set of eigenvectors to employ
in the formula (\ref{Bures}), and, thereby, can not further
 pursue the analysis.
Interestingly though, we have been able to determine the eigenvalues and
eigenvectors for the five-particle analogous scenario --- in which the single
parameter $\zeta_{1 1 1 1 0}$ must lie in the interval $[-{1 \over 7},
{1 \over 5}]$. The volume element of the Bures metric is, then,
\begin{equation}
{\sqrt{15} \sqrt{1 -21 \zeta_{1 1 1 1 0}^2} \over
2 \sqrt{-1 + 3 \zeta_{1 1 1 1 0}} \sqrt{1 +3 \zeta_{1 1 1 1 0}}
\sqrt{-1 +5 \zeta_{1 1 1 1 0}} \sqrt{1 +7 \zeta_{1 1 1 1 0}}}.
\end{equation}
We have been unable, however, to either normalize this and/or derive a
corresponding partition function.
\section{Quantum Canonical Ensemble for a Certain
 Three-Level Extension of the Two-Level Systems} \label{s5}
In our final analysis, we apply the same line of reasoning utilized in the
previous scenarios
  to recent results \cite{slat2}
concerning a particular {\it three}-level extension
 of the {\it two}-level systems.
These were given by density matrices of the form,
\begin{equation} \label{three}
\rho = {1 \over 2} \pmatrix{v+z & 0 & x- i y\cr
                              0 & 2 - 2 v & 0\cr
			x +iy &0&v-z\cr},
\end{equation}
so for $v=1$, the middle level is inaccessible and
 we recover the two-level systems.
The normalized volume element of the associated Bures metric has been found
to be \cite[eq. (17)]{slat2},
\begin{equation} \label{normprob}
{3 \over 4 \pi^{2} v \sqrt{1-v} \sqrt{v^2-x^2-y^2-z^2}}.
\end{equation}
From this, one can obtain (using spherical coordinates in the integrations),
 the univariate
marginal distribution (an asymmetric beta distribution) over the interval
$v \in [0,1]$ \cite[Fig. 3]{slat2},
\begin{equation} \label{marginal2}
{3 v \over 4 \sqrt{1-v}}.
\end{equation}
Interpreting this as the appropriate (normalized form of the)
 structure function, multiplying
by the Boltzmann factor $e^{-\beta h v}$ and integrating the product
over $v$ from 0 to 1, we obtain the corresponding partition function,
\begin{equation} \label{partthree}
Q(\beta) = {3 e^{-\beta h} ((1 +2 \beta h) \sqrt{\pi}
 \mbox{erfi}(\sqrt{\beta h})
- 2 \sqrt{\beta h} e^{\beta h}) \over 8 (\beta h)^{3/2}},
\end{equation}
from which the thermodynamic behavior of an ensemble of such systems
(\ref{three})
can be deduced. For instance, the expected value of the ``energy'' ($h v$)
 is given by
\begin{equation} \label{expv}
- {\partial \log{Q(\beta)} \over \partial \beta} =
{ (4 \beta^2 h^2 +4 \beta h +3) \sqrt{\pi} \mbox{erfi} (\sqrt{\beta h})
-2 e^{\beta h} \sqrt{\beta h} (2 \beta h +3) \over
2 \beta ((2 \beta h + 1) \sqrt{\pi} \mbox{erfi} (\sqrt{\beta h})
-2 e^{\beta h} \sqrt{\beta h})}.
\end{equation}
As $\beta \to 0$, this expected value approaches ${4 h \over 5}$, while the
variance about the expected value approaches ${8 h^2 \over 175}$.

\section{Analyses Based on the {\it Maximal}
  Monotone Metric} \label{parallel}

Subsequent to the distribution of their original preprint \cite{brodun},
D. C. Brody kindly informed me that he and his co-author, L. Hughston, had
``noticed that the phase space weighting used
 was incorrect, and as a result we find
a different expression for the expected energy. Thus, although the
claim of the paper is unchanged, we are now revising the example
considered therein. Interestingly, the same (what we now believe is the
correct) result was also predicted in our stochastic thermalisation
model (quant-ph/9711057) --- that is, instead of the ratio of the
Bessel functions, ${I_{2} \over I_{1}}$, we found the result given by the
Langevin function.''

In a second message (also prior to the appearance of \cite{brod}),
 Brody wrote: ``The only change [in \cite{brodun}]
we had to make is to replace the `phase space volume' by `state density'
(i. e. weighted volume). As a consequence, this results in changing the
calculation involving the spin-${1 \over 2}$ example. Nevertheless the
behaviour of the Langevin function is somewhat similar to that of the
ratio of the Bessel functions (e. g., energy, heat capacity).''

This reassessment of Brody and Hughston is paralled by the replacement
 of the {\it minimal} monotone metric by the
{\it maximal} monotone metric (of the left logarithmic derivative),
 since the use of the latter (coupled with a limiting argument), in fact, 
 leads to the Langevin function, that is the ratio
 ${I_{3/2} \over I_{1/2}}$, and not 
${I_{2} \over I_{1}}$.  (For a comprehensive treatment of
 monotone metrics and the ``geometries of quantum states'', see
\cite{petz}.)
Such a view would, in addition,
 be apparently consistent with an earlier study
of the author \cite{slat7}. There, it was found that the
maximal monotone metric yielded a more ``noninformative'' (that is, more
``neutral'' or less ``biased'') prior, for Bayesian analysis, than did
the minimal monotone metric (cf. \cite{petztoth}).
 (In the thermodynamic context, we form
posterior distributions by multiplying prior distributions
 by Boltzmann factors, while in the
Bayesian context, posteriors are formed by multiplying prior
distributions by the likelihoods
of a certain {\it finite} set of outcomes.)

Let us elaborate somewhat upon these points. The volume elements of both
the minimal and maximal monotone metrics over the three-dimensional convex
set (``Bloch sphere'') of spin-${1 \over 2}$ systems are proportional to
expressions of the form,
\begin{equation} \label{metrics}
{1 \over (1 - \xi_{1}^2 -\xi_{2}^2 -\xi_{3}^2)^u},
\end{equation}
where $u= {1 \over 2}$ in the minimal case, and ${3 \over 2}$ in the maximal
one \cite[eq. (3.17)]{petz}.
 In the minimal case, (\ref{metrics}) can be directly normalized
over the Bloch sphere $(\xi_{1}^2 +\xi_{2}^2 +\xi_{3}^2 \leq 1)$,
 and we are able to obtain univariate marginal probability
distributions of the type,
\begin{equation}
{2 \sqrt{1 -\xi_{1}^2} \over \pi}.
\end{equation}
Employing this result in
 Poisson's integral representation of the modified spherical
Bessel functions (cylinder functions of half integral order) \cite{slat1,kar},
 we recover
 the associated partition function (\ref{part}), as originally reported
by Brody and Hughston \cite{brodun}, and independently, Slater \cite{slat1}.
For $u={3 \over 2}$, (\ref{metrics}) is {\it not}
 normalizable (that is, improper)
over the Bloch
sphere. However, it can be  normalized over a three-dimensional ball of
radius $R <1$. Then, integrating out one of the three Cartesian coordinates
(say, $\xi_{3}$) over this ball of radius $R$, we obtain the bivariate
probability distribution,
\begin{equation}
{(R^2 -1) \sqrt{R^2 -\xi_{1}^2 -\xi_{2}^2} \over
2 \pi \sqrt{1-R^2} (-1 +\xi_{1}^2 +\xi_{2}^2) (R \sqrt{1 -R^2} + (R^2-1)
 \sin^{-1}R)}.
\end{equation}
In the limit $R \rightarrow 1$, this converges to the bivariate probability
distribution,
\begin{equation} \label{penult}
{1 \over 2 \pi \sqrt{1-\xi_{1}^2 -\xi_{2}^2}},
\end{equation}
over the unit disk ($\xi_{1}^2 + \xi_{2}^2 \leq 1$).
Its two univariate marginal probability distributions over $\xi_{1}$ and
$\xi_{2}$ are
simply uniform distributions $({1 \over 2})$ over the interval [-1,1].
 They, then,  give rise (again, applying Poisson's integral representation)
 to the Langevin function --- as in the
revision by Brody and Hughston (cf. sec.~\ref{s2}).
It is also of interest to note that if we {\it ab initio} set one of the
coordinates to zero in the volume element for the minimal
metric (\ref{metrics}), then the thermodynamic properties in such a
{\it  conditional}
case are identical to  those found
in the maximal monotone analysis. So, we have two independent ways of arriving
at the revised (and final) result of Brody and Hughston.

We have, in fact, conducted several analyses parallel to those reported
above in the main body of the paper, but
 based on the maximal, rather than minimal
 monotone metric. (To obtain the specific forms of the metric,
 we solved --- for the two-particle
scenarios ---  sets
 of thirty-two linear simultaneous equations \cite[eq. (4.26)]{helstrom}.)
 We take this opportunity
to report these results.

For the analysis of sec.~\ref{s1}, we arrived at a maximal monotone
 metric simply
proportional to the minimal monotone
metric ((\ref{BB})-(\ref{BBB})) and, hence, the same
 normalized density-of-states (\ref{marginal}) and partition
function (\ref{partcoupled}).

For the analysis of sec.~\ref{s2}, we  obtained a volume element of
the maximal monotone metric equal to
\begin{equation}
{\sqrt{2} \sqrt{\zeta_{11}^{2} +2 
{\xi_{1}^{(a)}}^{2} -1} \over \sqrt{1-\zeta_{11}^{2}}
(1- 4 {\xi_{1}^{(a)}}^{2} -\zeta_{11}^{2})}.
\end{equation}
Its
 integral over the domain of feasible values diverges, however, and we have not
been able to derive results analogous to (\ref{normvol}) and
(\ref{lang1}).

For the analysis of sec.~\ref{s2-3}, we arrive at
 precisely the same results as 
before,
that is,
(\ref{newprior}) and (\ref{new2}). Similarly, we obtain no differences
with the analysis of sec.~\ref{s3}.

For the three-particle analysis of sec.~\ref{s4}, the normalized volume
element of the maximal monotone metric is (cf. (\ref{unsucc})),
\begin{equation} \label{mm1}
{\sqrt{3} \over \pi \sqrt{1 -3 \zeta_{111}^2}},
\end{equation}
so that
\begin{equation} \label{mm2}
Q(\beta) = I_{0}({\beta h \over \sqrt{3}}).
\end{equation}
(In the minimal monotone or Bures metric analysis, on the other hand,
 we were unable to find
an explicit formula for the partition function.) For the four-particle
analysis of sec.~\ref{s4}, the analogous results based on the maximal
monotone metric are precisely the same
 ((\ref{four1}) - (\ref{four3})). The maximal-monotone-metric
analysis of the five-particle scenario yielded the same outcomes
((\ref{mm1})-(\ref{mm2})) as the
maximal-monotone-metric analysis of the three-particle scenario.

For the three-particle scenario of sec.~\ref{s10}, an analysis based on the
maximal monotone metric yielded the same results ((\ref{sd1})-(\ref{sd3})).
As before, we have not been able to compute the volume element for the
associated
four-particle scenario.

For the analysis of sec.~\ref{s5}, corresponding to a certain three-level
extension of the two-level systems, we obtain a volume
 element (cf. (\ref{normprob})),
\begin{equation}
{1 \over \sqrt{1-v} (v^2 -x^2-y^2-z^2)},
\end{equation}
{\it not} proportional to (\ref{normprob}). However,
 using spherical coordinates and a
 limiting argument (that is, integrating over the radial coordinate $r$
from 0 to ${v \over R}$, normalizing this result,
 and then letting $R \rightarrow 1$),
 it can be shown to yield precisely the same univariate
 marginal probability 
distribution (\ref{marginal2}) as for the analysis based on the Bures metric,
and hence lead to the same partition function (\ref{partthree}).

It would be of interest to attempt to find a characterization of those
cases in which the maximal and monotone metrics are the same (and so, of
course, are
their volume elements) or, more generally (as in the three-level
extension example), of the
cases
 in which the metrics may not
be identical, but the associated thermodynamic properties are,
nevertheless, the same.

Let us further note that for the {\it five}-dimensional convex set of
 {\it quaternionic} two-level quantum systems, the volume elements of the
maximal and minimal monotone metrics are proportional to expressions of the
form (cf. (\ref{metrics}), \cite[eqs. (7),(20)]{slat8}),
\begin{equation} \label{last}
{1 \over (1-\xi_{1}^2 -\xi_{2}^2 -\xi_{3}^2 -\xi_{4}^2 - \xi_{5}^2)^u},
\end{equation}
where $u = {1 \over 2}$ in the minimal case and ${5 \over 2}$ in the
maximal case. We then find --- again applying Poisson's integral
representation to the univariate marginal probability distributions obtained
from (\ref{last}) --- that in the minimal case, the expected energy is
proportional to the ratio ${I_{3} \over I_{2}}$ and in the maximal case
 (applying a similar limiting argument, as used to obtain (\ref{penult}))
to the ratio ${I_{5/2} \over I_{3/2}}$. So, these results are quite
analogous to those for the three-dimensional convex set of {\it complex}
two-level quantum systems.

Let us note that the results for the minimal and maximal monotone metrics
coincide for precisely those scenarios in which we are concerned
with a mutually {\it commuting} set of density matrices. Then, the scenario
is essentially {\it classical}, rather than quantum, in nature. Hence, the
monotone metric --- the one associated with the Fisher information --- is
simply
{\it unique}.
We will also point out that in a related analysis \cite{slaternew}, we
found the maximal monotone metric to be considerably more amenable to
our (MATHEMATICA) computations than was the minimal monotone metric.
\section{Concluding Remarks} \label{concluding}
As with the instance of a {\it single} spin-${1 \over 2}$ particle studied
in \cite{brodun,slat1} --- giving the previously reported results (\ref{part})
 and (\ref{energy}) ---  the possible applicability (to {\it small}
systems, in particular \cite{brod}) of the
several quantum canonical
ensembles presented in this letter, awaits experimental examination.
(We should note that it is usually considered that entangled particles
will decohere {\it before} they are thermalized,
that is, entanglement will be lost on time scales short compared
to those of thermal relaxation processes associated with energy exchange
with the bath \cite{mozyrsky}.)

Although we have examined a number of possible scenarios here, there is clearly
much opportunity for further systematic exploration along related lines.
In fact, in \cite{slaternew},  we were  able to assign a
prior probability distribution of the form ${15 (1-a) \sqrt{a}
\over 4 \pi \sqrt{b} \sqrt{c}}$ to the two-dimensional
simplex of diagonal entries --- $a, b, c$ --- of the eight-dimensional
convex set of spin-1 density matrices. (The evident symmetry between
these three entries had been broken by a particular transformation,
suggested by work of Bloore \cite{bloore}.) Then, we  obtained
a spin-1
 counterpart --- \cite[Fig. 4]{slaternew} --- to the Langevin function
(${I_{{3/2}} \over I_{{1/2}}}$).
These results pertained not to the {\it minimal} monotone metric
(which proved to be more computationally problematical),
but only to the {\it maximal} one --- which is the subject
 of sec.~\ref{parallel} --- and
 is consistent with the reassessment by Brody and Hughston  \cite{brod}
 of their
original work \cite{brodun}, as the maximal monotone metric applied
to a single spin-${1 \over 2}$ system, yields the Langevin function.
As noted above, this also accords with certain arguments of 
 Lavenda \cite[pp. 20, 193, 198]{lav},
who contends that the Langevin function has a sound probabilistic basis,
while 
the Brillouin function does not.
Also, as previously pointed out, the maximal monotone metric has been shown
in two separate analyses \cite{slat7,petztoth}
 to be more {\it noninformative} in character
than the minimal monotone metric. Other information-theoretic properties
of these metrics have been studied in \cite{kratt} (cf. \cite{slatbures}).

This study might be viewed as an attempt to respond
to the concluding admonition of Petz and Sud\'ar \cite[p. 2672]{petz}
that ``more than one privileged metric shows up in quantum mechanics.
The exact clarification of this point requires and is worth further studies''.
In particular, we have been concerned with evaluating the 
various thermodynamic
implications of these monotone metrics.

\acknowledgments

I would like to express appreciation to the Institute for Theoretical
Physics for computational support in this research.

\begin{center}
APPENDIX. Information Gains from Differing Possible
 Outcomes of Joint Spin Measurements
\end{center}
\appendix

We avail ourselves of the several prior probability distributions
(normalized volume elements of Bures metrics)
reported in sec.~\ref{minimal}, to obtain the information gains
 for various  outcomes of joint measurements of the partially
entangled spin-${1 \over 2}$ particles. (we might also pursue similar
analyses using the results of sec.~\ref{parallel}, based on
the maximal monotone metric.) We pass, in this manner, from
 thermodynamic considerations to ones of a fundamentally Bayesian nature. We
follow a methodology previously employed for {\it unentangled}
 spin-${1 \over 2}$
particles\cite{slat9}.

 We associate with each specific outcome
of a joint measurement of the partially entangled particles, a
likelihood function, the product of which with the prior distribution can be
normalized to yield --- {\it via} Bayes' rule --- a posterior probability
 distribution.
We then compute the information gain (Kullback-Leibler statistic) of
the posterior with respect to the prior. The products of the gains and
their corresponding likelihoods averaged over all the possible outcomes
 with
respect to the prior,
 then, give the {\it expected} information gain.

For the scenario of sec.~\ref{s1}, if we measure the spins of particles
$a$ and $b$ along the axis of polarization,
 and find them to disagree (that is, one spin ``up'' and
the other, ``down''), we gain
 $\log{3} - {2 \over 3}
\approx .431946$ nats of information, while if they agree (both spins either
up or down), we gain
considerably  less, that
is, $ \log{6} -{5 \over 3} \approx .125093$.
(The likelihood to be used in the Bayes' rule
 of obtaining a disagreement is ${(1-\zeta_{11}) \over 2}$,
while that for an agreement is ${(1+\zeta_{11}) \over 2}$. The  expected
information gain is .329662.)
 If after one 
agreement, we obtain a  disagreement, then the
 information gain due to
the second measurement is greater than that
 from an initial disagreement, that is,
$\log{5} - {16 \over 15} \approx .542771$. If, instead, there is a second
agreement, the additional information gain due to that outcome is only
.0427712 nats. Further, if an initial disagreement is followed by another, the
gain is $\log{{5 \over 3}} - {2 \over 5} \approx .110826$, while if the
initial
disagreement is succeeded by an agreement,
the gain is $ \log{10} -{31 \over 15} \approx  .235918$.

If we were to alter
 the scenario of sec.~\ref{s1}, setting $\xi_{1}^{(b)}$ equal
to the negative of $\xi_{1}^{(a)}$ rather than to $\xi_{1}^{(a)}$ itself (so
that the spins would be anticorrelated),
we anticipate obtaining the same expected information gains, but for the 
reversed situations.
For example,  the gain due to an initial
agreement would be .431946 nats, and not .125093, which would
be that associated with an initial disagreement, \ldots.

Contrastingly, for the scenario of sec.~\ref{s2},
 the information gain from an initial agreement is the same as that for
an initial 
disagreement, that  is $\log{2} -{1 \over 2} \approx .193147$, so this is
also the expected gain.
If a second measurement yields an outcome different from the first,
the information gain from this measurement is $\log{3} -{5 \over 6}
\approx .265279$, while if the outcome of the
 second measurement is the same
as the first, the gain is $\log{{3 \over 2}} - {1 \over 3}
\approx .0721318$ nats.

For the scenario of sec.~\ref{s3}, an initial agreement of spin
measurements of particles $a$ and $b$ along an arbitrary axis yields an
information gain of .306853 nats (saturating the Holevo-type bound
\cite{yuen,hall}), while an initial
 disagreement gives .0646381 nats. (The expected information gain is
.145376 nats.) A second agreement yields .0680544 nats,
while a second disagreement gives .0470689 nats.
An agreement after an initial disagreement results in a gain of
.427868 nats, while a disagreement after an initial agreement
provides .0516789 nats.

There, of course, still remains the possibility of using
positive-operator-valued measures (POVM's) and/or standard
measurements on {\it multiple} copies of (partially)
entangled systems of spin-${1 \over 2}$ particles \cite{slat9,pop,peres}.
Along similar lines,
in \cite{slat9} it was found (based on the corresponding Bures metric)
 that for a pair of identically prepared {\it unentangled} spin-${1 \over 2}$
particles,
 one could
expect to gain: (a) .357229 nats of information with the
 use of a particular POVM
 proposed by Peres \cite{peres} (having a continuum
of possible outcomes); (b) .313478
nats for a certain  joint measurement scheme of Popescu and Massar
\cite{pop} (cf. \cite{derkaa,latorre}),
 having four possible outcomes;  (c) .280372 nats
 if the spins of the two particles
were  to be measured separately in orthogonal directions; and (d) 
.212642 nats if
measured separately in the same direction.
A spin measurement on a 
\linebreak
{\it single} spin-${1 \over 2}$
 particle can be expected to yield
$.140186 = {.280372 \over 2}$ nats \cite{slat9}.

\end{document}